# Topotactic synthesis of a new BiS$_2$-based superconductor Bi$_2$(O,F)S$_2$

Tomoyuki Okada[1]*, Hiraku Ogino[1]*, Jun-ichi Shimoyama[1], and Kohji Kishio[1]

[1] *Department of Applied Chemistry, The University of Tokyo, 7-3-1 Hongo, Bunkyo-ku, Tokyo, 113-8656, Japan*

E-mail : 8781303601@mail.ecc.u-tokyo.ac.jp

A new BiS$_2$-based superconductor Bi$_2$(O,F)S$_2$ was discovered. This is a layered compound consisting of alternate stacking structure of rock-salt-type BiS$_2$ superconducting layer and fluorite-type Bi(O,F) blocking layer. Bi$_2$(O,F)S$_2$ was obtained as the main phase by topotactic fluorination of undoped Bi$_2$OS$_2$ using XeF$_2$, which is the first topotactic synthesis of an electron-doped superconductor via reductive fluorination. With increasing F-content, *a*- and *c*-axis length increased and decreased, respectively, and $T_c$ increased up to 5.1 K.

Since the discovery of superconductivity in LaFeAs(O,F) with superconducting transition temperature ($T_c$) of 26 K[1], superconductivity in novel layered compounds have attracted intense attention. Consequently, several new layered superconductors with Ti$_2$O plane (ex. Ba$_2$Ti$_2$Sb$_2$O with $T_c$ = 1.2 K[2]), and BiS$_2$ layer (ex. Bi$_4$O$_4$S$_3$ with $T_c$ = 8.3 K[3]) are recently discovered. Some of these newly discovered superconductors, such as iron-based superconductors $RE$FeAs(O,F)[1] and BiS$_2$-based superconductors $RE$(O,F)BiS$_2$ ($RE$ = La[4], Ce[5], Pr[6] and Nd[7]), contains fluorite-type $RE$(O,F) blocking layers. In these cases, undoped $RE$FeAsO and $RE$OBiS$_2$ does not show superconductivity and partial substitution of F$^-$ for the O$^{2-}$ site causes electron-doped superconductivity in the FeAs and BiS$_2$ layer.

Not only rare earth metals, but also Bi forms fluorite-type BiO layer. Among the BiS$_2$-based superconductors reported to date, Bi$_4$O$_4$S$_3$[3] and Bi$_3$O$_2$S$_3$[8)9)] contains fluorite-type BiO layer in the blocking layer. In addition, Bi$_2$OS$_2$ (i.e. BiOBiS$_2$) is isostructural with $RE$OBiS$_2$, consisting of rock-salt-type BiS$_2$ layer and fluorite-type BiO layer, as shown in **Fig.1**. This compound was synthesized first by Phelan *et al*. in 2013 and reported to be a non-superconductor[8]. However, no further investigation on physical properties of this compound has been reported, while partial substitution of F$^-$ for the O$^{2-}$ site of fluorite-type BiO layer might lead to superconductivity.

In general, conventional solid-state reaction has been usually adopted for introduction of fluorine to oxides, where metal fluorides and other raw or precursor materials are mixed at the stoichiometric composition of the objective substance and reacted at high temperatures. On the other hand, various topotactic fluorination reactions have also been reported to date, mainly on transition metal oxides having perovskite-type structures, where precursor oxides reacts on fluorinating agents, such as F$_2$ gas[10], XeF$_2$[11),12)], NH$_4$F[13], $M$F$_2$ ($M$ = Ni, Cu, Zn, Ag)[14], poly (vinylidene fluoride) (PVDF)[15], and polytetrafluoroethylene (PTFE)[16]. When F$^-$ is inserted to the interstitial site of the transition metal oxide, oxidation state of the transition metal increases. This type of reaction; topotactic oxidative fluorination, is sometimes adopted to prepare hole-doped high-$T_c$ cuprates, such as Sr$_2$CuO$_2$F$_{2+\delta}$[10), 12)-14)]. On the contrary, when F$^-$ is partially substituted for the O$^{2-}$ site, the oxidation state of transition metal decreases. Synthesis of several oxyfluorides, such as CrOF$_3$[11], RbLaNb$_2$O$_6$F[16] and Sr$_3$Fe$_2$(O,F)$_7$[17], were reported via this route, *i.e.* topotactic reductive fluorination, but any superconductors have never been discovered through this method.

In the present Letter, synthesis and properties of F-doped Bi$_2$OS$_2$ are reported. Bi$_2$(O,F)S$_2$ was successfully synthesized by topotactic reaction between undoped Bi$_2$OS$_2$ and XeF$_2$, and showed superconductivity with $T_c$ = 5.1 K. This is the first topotactic reductive fluorination resulting in an electron-doped superconductor.

Undoped $Bi_2OS_2$ was synthesized by the solid-state reaction. $Bi_2O_3$ (Furuuchi Chemical, 99.9%) and $Bi_2S_3$ (ALDRICH, 99%) powders were weighed at the stoichiometric composition of $Bi_2OS_2$, mixed, pelletized, sealed in an evacuated quartz ampoule and heated at 300~600°C with several times of regrinding process. $Bi_2(O,F)S_2$ was prepared by two different methods. The first procedure is simple solid-state reaction. Powders of $Bi_2O_3$, $Bi_2S_3$, $BiF_3$ (Kojundo Chemical Laboratory, 99.9%) and Bi (Furuuchi Chemical, 99.9%) were mixed at the nominal compositions of $Bi_2(O_{1-x}F_x)S_2$ and heated at 300 ~ 500°C for 12 ~ 100 h with intermediate grindings. The second procedure is the topotactic reaction between undoped $Bi_2OS_2$ and $XeF_2$. Undoped $Bi_2OS_2$ synthesized by solid-state reaction and $XeF_2$ (Matrix Scientific, 99%) powder were weighed, mixed, pelletized, sealed in an evacuated quartz ampoule and heated at 100 ~ 450°C for 24 ~ 72 h to form $Bi_2(O,F)S_2$. Note that during the mixing and pelletizing process, the mixture easily causes explosion thus careful treatment under Ar-filled glove box was mandatory.

The constituent phases of the samples were analyzed by X-ray diffraction (XRD) measurement (RIGAKU Ultima-IV) with Cu-$K_\alpha$ radiation generated at 40 kV and 40 mA. Silicon powder was used as the internal standard to determine the lattice parameters. Measurements of magnetization and resistivity were performed by a SQUID magnetometer (Quantum Design MPMS-XL5s) and by the AC four-probe method using the Quantum Design PPMS, respectively. The thermoelectric power was measured by the steady-state method.

Undoped $Bi_2OS_2$ polycrystalline bulks with almost single phase were obtained by sintering at 300°C for ~100 h with two or three times of intermediate grindings as shown in **Fig. 2**. The diffraction peaks were well indexed to the space group *P4/nmm*, and the lattice constants were calculated to be $a$ = 3.961 Å and $c$ = 13.808 Å. Small amount of impurity phases, such as $Bi_2S_3$ and Bi, always remained in this study.

$Bi_2(O_{1-x}F_x)S_2$ was at first synthesized by the conventional solid-state reaction using $BiF_3$ as a fluorine source. For a sample with $x$ = 0.3, $Bi_2(O,F)S_2$ formed with a lot of impurity phase of BiOF, $Bi_2S_3$ and Bi as shown in **Fig. 2**. Lattice constants were calculated to be $a$ = 3.969 Å and $c$ = 13.769 Å, which are different from those of undoped $Bi_2OS_2$, suggesting partial substitution of $F^-$ for the $O^{2-}$ site. This sample showed large diamagnetic signal below ~4 K in the zero-field-cooled (ZFC) magnetization measurement under 1 Oe. When $x$ is increased to 0.5, the sample contained only BiOF, $Bi_2S_3$ and Bi. This sample and undoped $Bi_2OS_2$ sample did not show large diamagnetism above 2 K, while very small diamagnetism without temperature dependence was observed possibly due to the diamagnetism of Bi, which is not related to superconductivity.

In order to obtain $Bi_2(O,F)S_2$ with higher purity and evaluate more intrinsic superconducting properties, we employed different synthesis route; topotactic reaction between undoped $Bi_2OS_2$ and $XeF_2$. We expected that reductive fluorination occurs in the present case according to the following chemical reaction:

$Bi_2OS_2$ (s) + ($x$/2) $XeF_2$ (s) → $Bi_2(O_{1-x}F_x)S_2$ (s) + ($x$/2) Xe (g) + ($x$/2) $O_2$ (g).     (1)

**Fig. 2** shows the XRD patterns of the samples synthesized via the topotactic reaction at 400°C for 72 h. $Bi_2(O,F)S_2$ is obtained as the main phase and amounts of co-existing impurity phases were largely reduced compared to the samples synthesized via conventional solid-state reaction. With an increase in nominal $XeF_2$ content up to ~30 mol% ($x$/2 = 0.3) of the $Bi_2OS_2$, the $a$-axis length slightly increased while the $c$-axis length largely decreased. This systematic change of lattice parameter indicates that the fluorine content is successfully varied by changing the nominal $XeF_2$ content, though investigation of actual fluorine content is difficult due to small difference in X-ray scattering factors between O and F. The actual fluorine concentrations in the samples are considered to be lower than the nominal $x$, because part of $XeF_2$ is considered to decompose into gases of Xe and $F_2$ before reacting with $Bi_2OS_2$. Nevertheless, we are sure that partial substitution of $F^-$ for the $O^{2-}$ site occurred in this case, following the reaction equation (1), and oxidation state of Bi is reduced by this reaction. One reason is the systematic changes of lattice parameters as described above. Both topotactic fluorination and fluorine doping via conventional solid-state reaction caused similar change of lattice parameters: expansion of $a$-axis and contraction of $c$-axis lengths. Large contraction of $c$-axis length by $F^-$ doping is consistent with the case of other $RE$(O,F)$BiS_2$ electron-doped superconductors[4)-7)], suggesting that electron doping is successfully achieved in $Bi_2OS_2$ by topotactic fluorination. The results of thermoelectric power measurement shown in **Fig. 3** strongly support this idea. Seebeck coefficient of both undoped and topotactically fluorinated $Bi_2OS_2$ are negative in the temperature range of 4.2 to 300 K and the absolute value of the thermoelectric power significantly decreased in $Bi_2(O,F)S_2$. At 300 K, Seebeck coefficient of undoped $Bi_2OS_2$ is ~70 μVK$^{-1}$, while that of $Bi_2(O,F)S_2$ fluorinated by $XeF_2$ with 50 mol% of $Bi_2OS_2$ is ~30 μVK$^{-1}$, indicating increase of electron type charge carriers.

**Fig. 4(a)** shows ZFC and field-cooled (FC) magnetization curves of $Bi_2(O,F)S_2$ samples with $x$/2 = 0.1 and 0.3 prepared by topotactic fluorination. Since both samples exhibited very sharp transition and perfect diamagnetism, $Bi_2(O,F)S_2$ is concluded to be the bulk superconductor. $T_c$ increases up to 5.1 K as increasing nominal $XeF_2$ content up to ~30 mol% of the undoped $Bi_2OS_2$. **Fig. 4(b)** shows the temperature dependence of resistivity measured for a sample with $x/2 = 0.3$. Very sharp superconducting transition within 0.4 K and zero resistivity at 4.8 K is observed under 0 T, while superconductivity is rapidly suppressed in magnetic fields and resistivity drop were not observed above 2 K under 3 T. The behavior of normal state resistivity seems metallic, contrary to the semiconducting behavior observed in polycrystalline samples of $RE(O,F)BiS_2$[4)-7)]. However, our sample contains small impurity of Bi metal, which might affect the normal state resistivity. Therefore, intrinsic behavior of normal state transport of this compound remains unrevealed.

Even when nominal $XeF_2$ content was increased up to ~50 mol% of $Bi_2OS_2$, the result was almost the same as the case when nominal $XeF_2$ content was ~30 mol%, probably owing to solubility limit of $F^-$ for the $O^{2-}$ site. Changing reaction temperature did not lead to increase of fluorine content. The sample reacted at 200~300°C also showed superconductivity below 4 ~ 5 K. Their superconducting transitions were broader with smaller superconducting volume fraction at ~2 K compared to the samples reacted for 400°C, suggesting insufficient substitution of fluorine. The sample reacted at lower than 200°C did not show apparent diamagnetism due to superconductivity. On the other hand, the reaction at 450°C lead to the decomposition of $Bi_2OS_2$ phase.

The $a$-axis length of $Bi_2(O,F)S_2$ is shorter than other isostructural $RE(O,F)BiS_2$[4)-7)] and $(Sr,La)FBiS_2$[18)]. In $RE(O,F)BiS_2$, $a$-axis shrinks and $T_c$ increases with increasing atomic number of $RE$ from La to Nd[4)-7)]. The $T_c$ of $Bi_2(O,F)S_2$ (5.1 K) is higher than those of $RE(O,F)BiS_2$ with $RE$ = La ~ Pr and comparable to that of $Nd(O,F)BiS_2$ (5.2 K) [7)]. It seems that smaller $a$-axis length is preferable for higher $T_c$ in $MOBiS_2$ synthesized under ambient pressure. However, $La(O,F)BiS_2$ shows large increase of $T_c$ up to ~10 K by annealing under high-pressure[4)]. In order to discuss the intrinsic relationship between lattice parameter and $T_c$, high-pressure studies are necessary.

Synthesis of $Bi_2(O,F)S_2$ using $XeF_2$ is the first example for topotactic synthesis of electron-doped superconductor via reductive fluorination process, while topotactic oxidative fluorination has been used to obtain hole-doped cuprates[10), 12)-14)]. We think that topotactic reductive fluorination can also be a powerful method to explore other novel superconductors. Besides, researches on topotactic fluorination[10), 12)-17)] and hydrogenation[19)20)] have been mainly focused on simple and layered perovskite-related materials, such as Ruddlesden-Popper type layered perovskite oxides, but our result indicate that topotactic reaction can also be applied on layered compounds containing fluorite-type layer.

In summary, a new member of BiS$_2$-based superconductor, Bi$_2$(O,F)S$_2$ with $T_c$ = 5.1 K was discovered. Topotactic reductive fluorination using XeF$_2$ in the temperature range of 200 ~ 400°C was found to be useful for obtaining Bi$_2$(O,F)S$_2$ as the main phase and the reaction at 400°C is suitable for successful control of fluorine content resulting in high quality superconductors with sharp transitions and large superconducting volume fractions. Our synthesis of Bi$_2$(O,F)S$_2$ using XeF$_2$ is the first topotactic reductive fluorination to obtain an electron-doped superconductor. Topotactic fluorination could be widely used for the control of physical properties in various compounds, including layered compounds with complex stacking structure, and this would be helpful for further exploration of novel superconductors.

**Acknowledgement**
This work was partly supported by the JSPS KAKENHI Grant Number 26390045, and Izumi Science and Technology Foundation. We thank to Dr. T. Fujii and Dr. R. Toda in the Cryogenic Research Center of the University of Tokyo for thermoelectric power and resistivity measurements. We also thank to Dr. A. Iyo, Dr. N. Takeshita and Dr. H. Eisaki in National Institute of Advanced Industrial Science and Technology for fruitful discussions.

*Note added* : During editorial process J. Shao *et al.* have reported superconductivity in Bi$_2$(O,F)S$_2$[21], while their sample synthesis method and $T_c$ (3.5 K) are different from ours.

# References

1) Y. Kamihara, T. Watanabe, M. Hirano and H. Hosono, J. Am. Chem. Soc., 130 (11), 3296-3297 (2008)
2) T. Yajima, K. Nakano, F. Takeiri, T. Ono, Y. Hosokoshi, Y. Matsusita, J. Hester, Y. Kobayashi, H. Kageyama, J. Phys. Soc. Jpn. 81 103706 (2012)
3) Y. Mizuguchi, H. Fujihisa, Y. Gotoh, K. Suzuki, H. Usui, K. Kuroki, S. Demura, Y. Takano, H. Izawa and O. Miura, Phys. Rev. B 86 220510(R) (2012)
4) Y. Mizuguchi, S. Demura, K. Deguchi, Y. Takano, H. Fujihisa, Y. Gotoh, H. Izawa and O. Miura, J. Phys. Soc. Jpn 81 114725 (2012)
5) J. Xing, S. Li, X. Ding, H. Yang and H.H. Wen, Phys. Rev. B 86 214518 (2012)
6) R. Jha, A. Kumar, S. K .Singh and V. P. S.Awana, J. Sup. and Novel. Mag. 29 499-502 (2013)
7) S. Demura, Y. Mizugichi K. Deguchi, H. Okazaki, H. Hara, T. Watanabe, S. J. Denholme, M. Fujioka, T. Ozaki, H. Fujihisa, Y. Gotoh, O. Miura, T. Yamaguchi, H. Takeya and Y. Takano, J. Phys. Soc. Jpn 82 033708 (2013)
8) W. A. Phelan, D. C. Wallace, K. E. Arpino, J. R. Neilson, K. J. Livi, C. R. Seabourne, A. J. Scott, and T. M. McQueen, J. Am. Chem. Soc., 135 (14), 5372-5374 (2013)
9) L. Li, D. Parker, P. Babkevich, L. Yang, H. M. Ronnow, and A. S. Sefat, arXiv : 1412.3070
10) M. Al-Mamouri, P. P. Edwards, C. Greaves and M. Slaski, Nature 369, 382-384 (1994)
11) M. McHughes, R. D. Willett, H. B. Davis and G. L. Gard, Inorg. Chem. 25 426-427 (1986)
12) E. I. Ardashnikova, S. V. Lubarsky, D. I. Denisenko, R. V. Shpanchenko, E. V. Antipov, G. V. Tendeloo, Physica C 338 259-265 (1995)
13) P. R. Slater, P. P. Edwards, C. Greaves, I. Gameson, J. P. Hodges, M. G. Francesconi, M. Al-Mamouri and M. Slaski, Physica C 241 151 (1995)
14) P. R. Slater, J. P. Hodges, M.G. Francesconi, P. P. Edwards, C. Greaves, I. Gameson, and M. Slaski, Physica C 253 16-22 (1995)
15) P. R. Slater, J. Fluorine Chem. 117 43 (2002)
16) Y. Kobayashi, M. Tian, M. Eguchi and T. E. Mallouk, J. Am. Chem. Soc., 131, 9849-9855 (2009)
17) Y. Tsujimoto, K. Yamaura, N. Hayashi, K. Kodama, N. Igawa, Y. Matsushita, Y. Katsuya, Y. Shirako, M. Akaogi, and E. Takayama-Muromachi, Chem. Mater. 23 3652-3658 (2011)
18) Y. Li, X. Lin, L. Li, N. Zhou, X. Xu, C. Cao, J. Dai, L. Zhang, Y. Luo, W. Jiao, Q. Tao, G. Cao and Z. Xu, Supercond. Sci. Technol, 27 035009 (2014)
19) M. A. Hayward, E. J. Cussen, J. B. Claridge, M. Bieringer, M. J. Rosseinsky, C. J. Kiely, S. J. Blundell, I. M. Marshall, F. L. Pratt, Science 295 1882-1884 (2002)
20) Y. Kobayashi, O. J. Hernandez, T. Sakaguchi, T. Yajima, T. Roisnel, Y. Tsujimoto, M. Morita, Y. Noda, Y. Mogami, A. Kitada, M. Ohkura, S. Hosokawa, Z. Li, K. Hayashi, Y. Kusano, J. Kim, N. Tsuji, A. Fujiwara, Y. Matsushita, K. Yoshimura,

K. Takegoshi, M. Inoue, M. Takano, and H. Kageyama, Nature Materials 11 507-511 (2012)

21) J. Shao, X. Yao, Z. Liu, L. Pi, S. Tan, C. Zhang, and Y. Zhang, Supercond. Sci. Technol. 28 015008 (2015)

**Figure Captions**

**Fig.1** Crystal structure of $Bi_2(O,F)S_2$. The unit cell is depicted with the solid line.

**Fig. 2** Powder XRD patterns of undoped and F-doped $Bi_2OS_2$ synthesized via solid-state reaction (the lower two lines) and $Bi_2(O,F)S_2$ synthesized via topotactic reaction with $XeF_2$ (the upper two lines).

**Fig. 3** Temperature dependence of Seebeck coefficient of $Bi_2(O,F)S_2$ synthesized via topotactic reaction with $XeF_2$ and undoped $Bi_2OS_2$.

**Fig. 4** Temperature dependence of **(a)** magnetization ($H$ = 1 Oe) and **(b)** resistivity of $Bi_2(O,F)S_2$ synthesized via topotactic reaction with $XeF_2$.

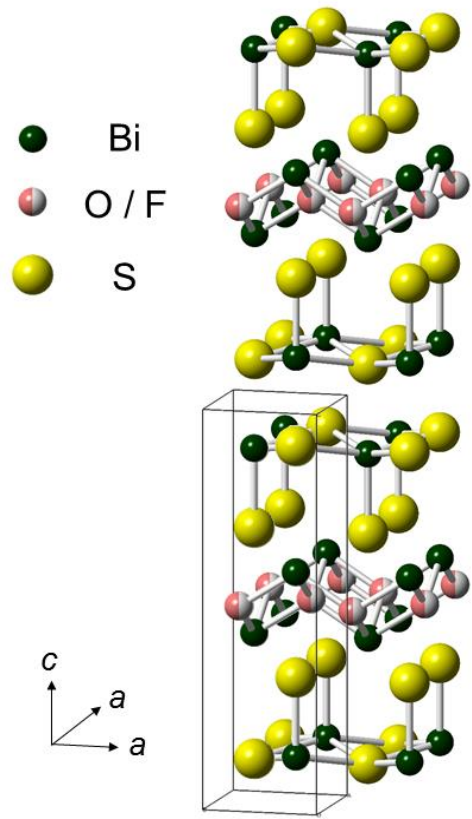

**Fig. 1.**

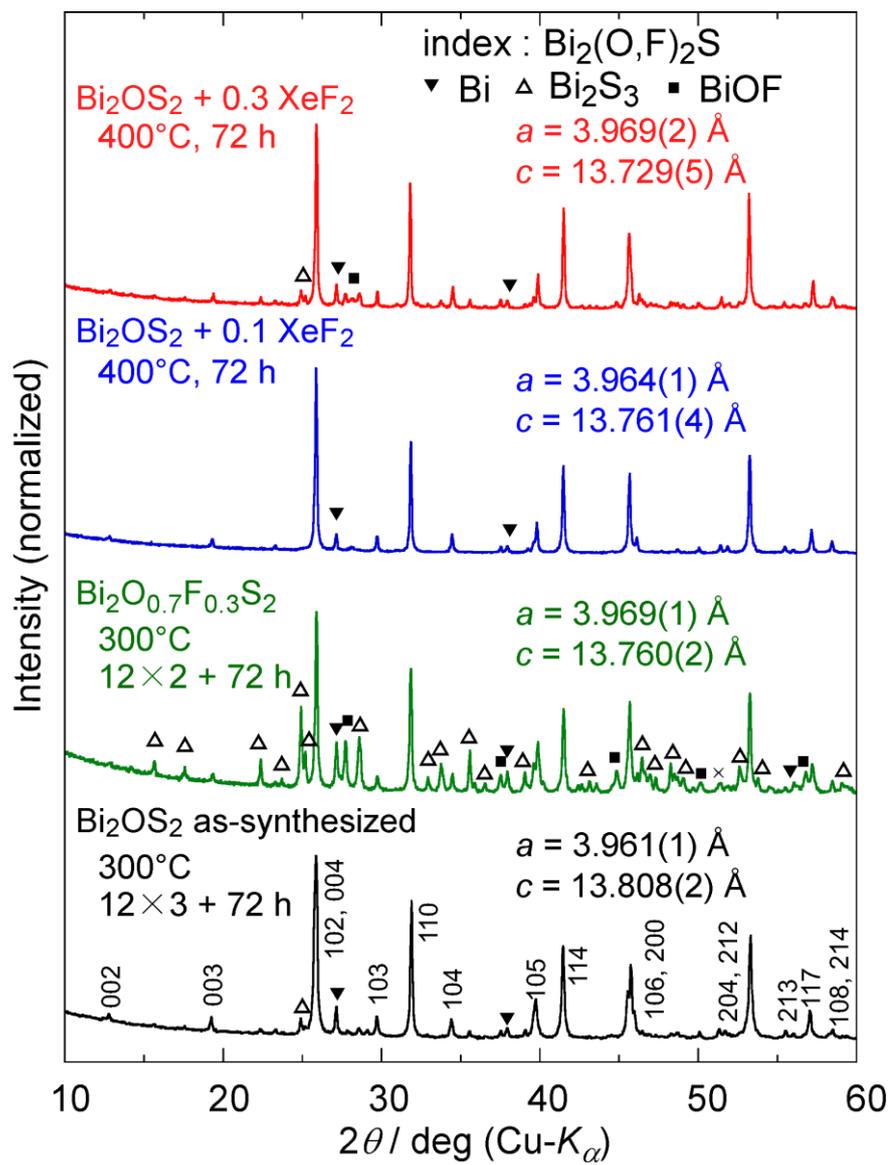

**Fig. 2.**

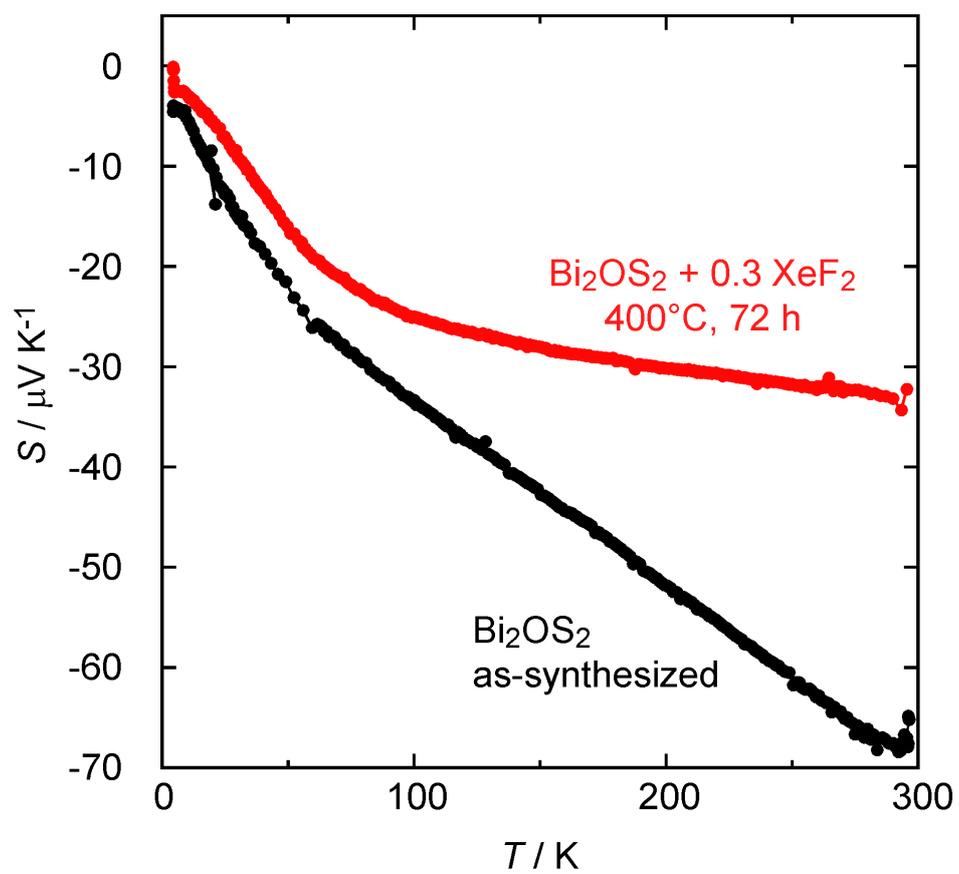

Fig. 3.

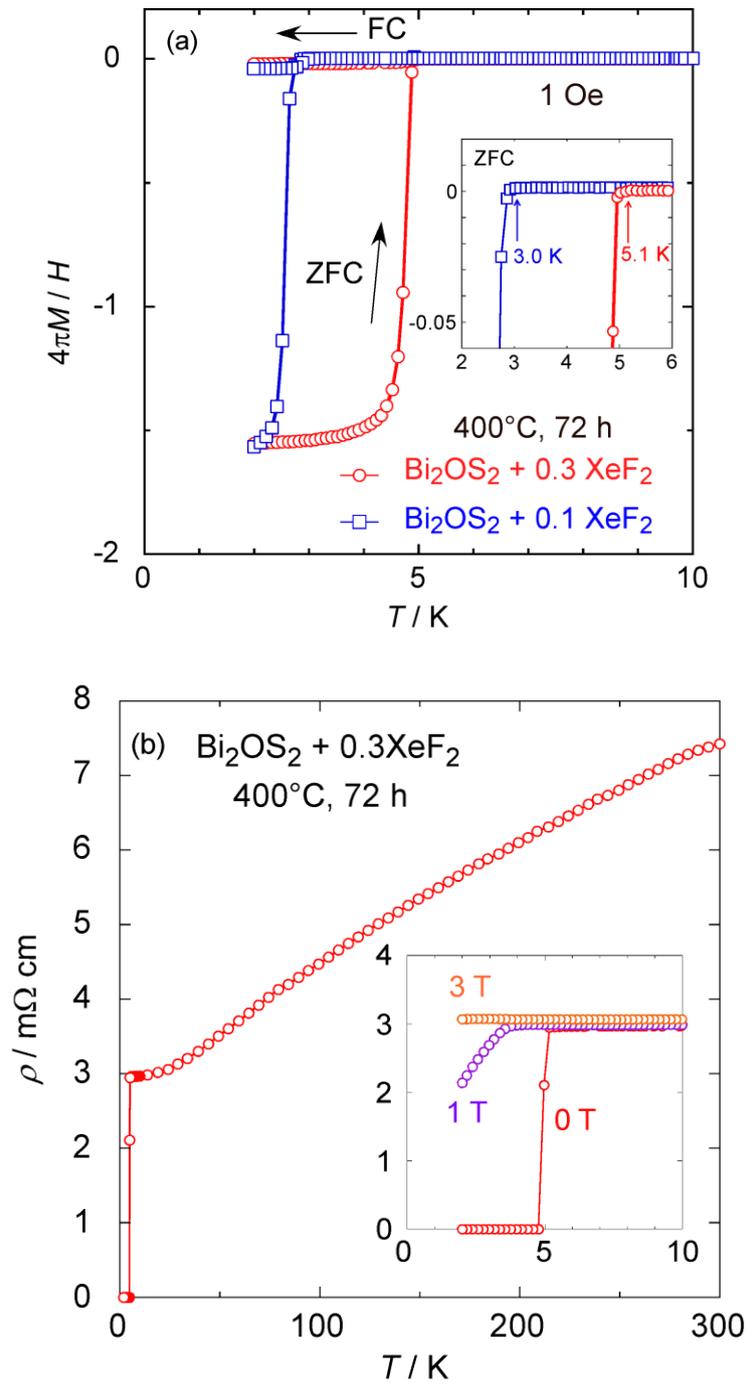

**Fig. 4.**